# Intelligent Resource Allocation Technique For Desktop-as-a-Service in Cloud Environment

Gandhi Kishan Bipinchandra,
M.Tech(CE) Research Scholar,
RK University, India,
Email:
gandhikishan13@gmail.com.

Prof. Rajanikanth Aluvalu ,
Dept. of Computer Engg,
RK University, India,
Email:
rajanikanth.aluvalu@rku.ac.in

Dr.Ajay Shanker Singh ,
Email:
drajay.cse@gmail.com.

**ABSTRACT: -**

*The specialty of desktop-as-a-service cloud computing is that user can access their desktop and can execute applications in virtual desktops on remote servers. Resource management and resource utilization are most significant in the area of desktop-as-a-service, cloud computing; however, handling a large amount of clients in the most efficient manner poses important challenges. Especially deciding how many clients to handle on one server, and where to execute the user applications at each time is important. This is because we have to ensure maximum resource utilization along with user data confidentiality, customer satisfaction, scalability, minimum Service level agreement (SLA) violation etc. Assigning too many users to one server leads to customer dissatisfaction, while assigning too little leads to higher investments costs. So we have taken into consideration these two situations also. We study different aspects to optimize the resource usage and customer satisfaction. Here in this paper We proposed Intelligent Resource Allocation (IRA) Technique which assures the above mentioned parameters like minimum SLA violation. For this, priorities are assigned to user requests based on their SLA Factors in order to maintain their confidentiality. The results of the paper indicate that by applying IRA Technique to the already existing overbooking mechanism will improve the performance of the system with significant reduction in SLA violation.*

## I. INTRODUCTION

Cloud computing refers technology that facilitate functionality of an IT Infrastructure, IT platform or an IT product to be exposed as a set of services in a seamlessly scalable model so that the consumers of these services can use what they really want and pay for only those services that they use (Pay per use).Cloud computing is about moving services, computation or data–for cost and business advantage-off–site to an internal or external, location transparent, centralized facility. By making data available in the cloud, it can be more easily and ubiquitously accessed, often at much lower cost, increasing its value by enabling opportunities for enhanced collaboration, integration and analysis on a shared common platform. The definition of cloud computing as per Gartner is "A style of computing where massively scalable IT facilitate capabilities are delivered as a service to external customers using internal technologies".

The Cloud computing services such as Amazon's Elastic are widely available today, offering computing resources on demand. Thanks to such advances and ubiquitous network availability, the thin client computing paradigm is enjoying increasing popularity. Originally intended for wired LAN environments this paradigm is repeating its success in a mobile context. A study from ABI Research forecasts a US$20 billion turnover surrounding services directly associated with mobile cloud computing by the end of 2014. Clearly, when applications are offloaded, the mobile terminal only needs to present audiovisual output to users and convey user input to remote servers, considerably reducing the client device's computational complexity. Consequently, applications can run as-is, without requiring (many) scaled-down versions for mobile devices[1].

Several popular applications, such as Google Docs and Microsoft Live, already execute on servers in the cloud. The ability to access applications in the cloud is referred to as software as a service (SaaS).There has been a rapid adoption of "cloud" platforms for online applications such as email, photo/video galleries and file storage in academia and industry. The next frontier for these user communities will be to transition their "traditional distributed desktops" that have dedicated hardware and software installations into "virtual desktop clouds" (VDCs) that are accessible via thin-clients. Moreover, in the not so distant future, we can envisage home users signing-up for virtual desktops (VDs) with a VD Cloud Service Provider (CSP) providing Desktop-as-a-Service (DaaS) as a utility[1].

Current desktop-as-a-service computing deployments are typically operational in corporate local area network (LAN) environments, which are highly controlled environments offering fixed and stable bandwidth availability to a relative small, well-known user base. Extending desktop-as-a-service computing to wide area network (WAN) environments, which comprise a large, geographically distributed customer base, where users are potential.

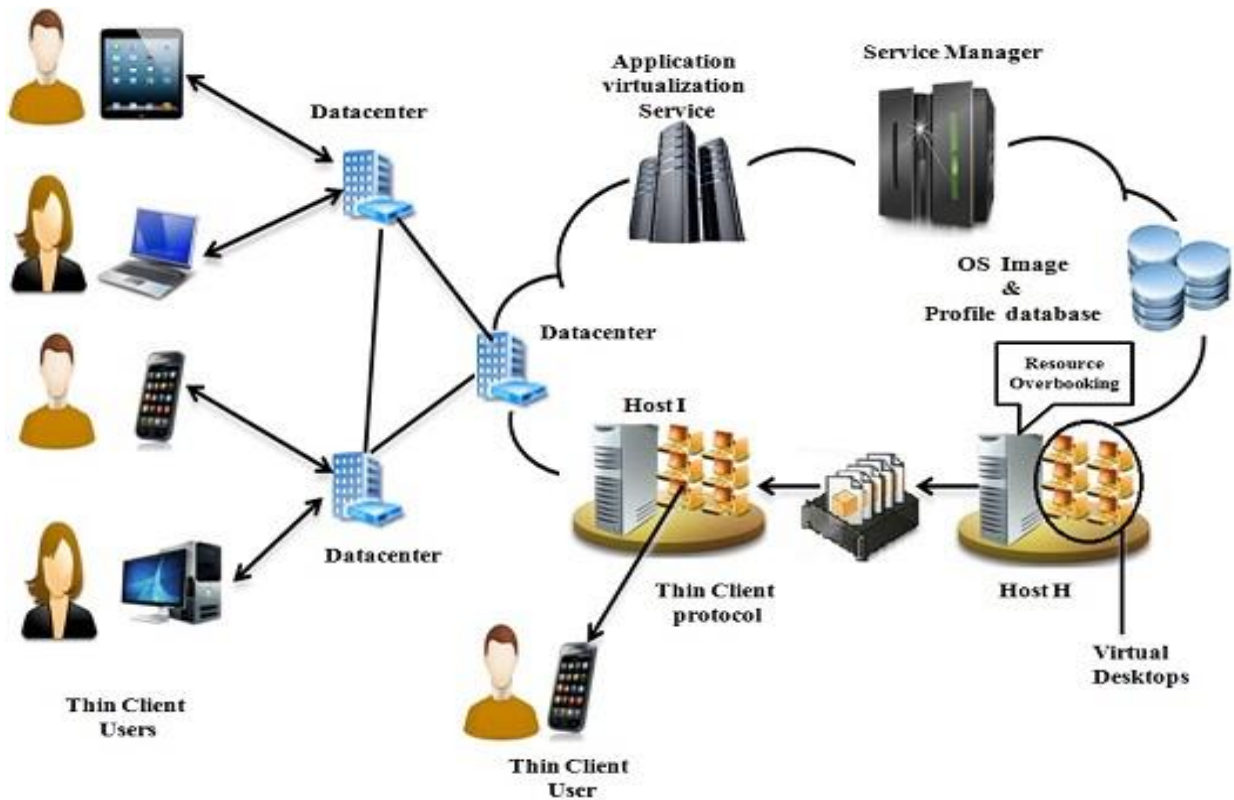

*Figure 1. System architecture for enabling cloud-based desktop services for thin clients. Users connect via a thin client device — a smart phone, tablet PC, PDA, netbook, or minimal- or zero-state device — to their remote applications executed in a virtual desktop. The service manager's self management component covers optimizations to improve the user experience and decrease service provider costs.*

connected through unreliable wireless network connections, involves a number of novel challenges. Strategies are needed to improve resource utilization and/or customer satisfaction in WAN environments in the most efficient manner. Cloud computing [2] is an enabler for this kind of service. Unlike most current cloud services, the applications are not accessed through a web browser (e.g. Google Apps Cloud Service), but through a thin client protocol (e.g., Microsoft Remote Desktop Protocol (RDP) [3] or Virtual Network Computing (VNC) [4]). This way, legacy applications must not be rebuilt to be offered by the envisioned service.

Current cloud platforms fulfill the hardware requirements for implementing DaaS. However, an emerging category of mobile applications including augmented reality, rich sensing, and multimedia editing pose stringent requirements on delays. Current cloud management systems can't meet user expectations for these applications, especially in terms of latency. A clear need exists for novel cloud management algorithms that consider the specific requirements of thin client computing. system architecture implements such algorithms in the service manager's self-management component. The manager can be implemented as part of existing cloud management systems such as OpenNebula , OpenStack, and Eucalyptus.

Figure 1 shows Existing architecture. Simplified OS image management (that is, re-using an OS image among users to reduce the storage per user) and application management are essential for the service to scale. Our system builds a VD from a shared golden image from the OS database and augments it with personal settings for example, by using a copy-on-write solution with UnionFS (http://unionfs.filesystems.org). Multilayer VDs simplify the complexity of upgrading the golden image without causing broken dependencies or conflicts.To improve DaaS usability, we could combine DaaS with application virtualization technologies such as Softricity and Microsoft App-V. The system would then dynamically deliver applications to the user's VD without having to install, configure, and update them. This approach further reduces the complexity of upgrading golden OS images because applications aren't installed in the user's VD and thus can't be broken[1].

## II. RELATED WORK

In desktop-as-a-service cloud computing, user applications are executed in Virtual Desktop on remote servers. This offers great advantages in terms of usability and resource utilization; however, handling a large amount of clients in the most efficient manner poses important challenges. Especially deciding how many clients to handle on one server, and where to execute the user applications at each time is important. Assigning too many users to one server leads to customer dissatisfaction, while assigning too little leads To higher investment costs of Service provider.

In cloud computing the resource management has got a cogent role in the performance of the whole process and the level of customer satisfaction provided by the process. But while providing the maximum customer satisfaction the service provider ought to be definite the profits that incur to them also. So the resource management ought to be efficient on both perspectives i.e. on the end user and the service provider point of view.

Here in this section we are going to analyze various the resource management strategies that are previously present in the cloud environment and their basic principle its positive as well as negative aspects.

In [4] it proposes a new autonomic workload provisioning that addresses the challenges of enterprise grids and clouds. The main aim of this mechanism is that to improve the resource utilization and which is achieved with the help of reducing the over provisioning. This paper also explored the use of workload modeling techniques and their application. The mechanism for dynamic and decentralized VM provisioning monitors the flow of arriving jobs from different queues in a decentralized manner during ongoing analysis windows of duration in the order of the startup time of new VMs. The cloud environment has to take into consideration all these things for each of its clients and could provide maximum service to all of its clients.

In [5] it suggests that when we are taking the scheduling of resources and tasks separately it imposes large waiting time and response time. In order to overcome this drawback a new approach namely Linear Scheduling for Tasks and Resources (LSTR) is introduced. Here scheduling algorithms mainly focus on the division of the resources among the requestors which will make the most of the selected QoS parameters. The QoS parameter selected in this approach is the cost function. The scheduling algorithm is designed based on the tasks and the available virtual machines together and named LSTR scheduling strategy. This is planned in order to boost utilization of resources.

In [6], it talks about the live migration of the virtual machines. In this paper they suggest that migrating the operating system instances across distinct physical hosts is a useful tool for the administrator of data centers and clusters. It also provides a separation between hardware and software and provides fault management, low level system maintenance and load balancing. Here an approach namely "pre-copy approach" is introduced. In this approach pages of memory are iteratively copied from the source machine to the destination host and in addition there is a fact that all these things are done without ever stopping the execution of the system. Page level protection hardware is used to make sure that a consistent snapshot is transferred. For controlling the traffic of other running services a rate-adaptive algorithm is used. And during the final phase it pauses the virtual machine and copies any remaining pages to the destination and after that resumes the execution there. The factors affecting the total migration are link bandwidth, migration overhead and page dirtied rate [7].

Roy et al. [8] describes about the cost based workload provisioning and "just- in- time resource allocation". Workload Prediction is the prediction of the workload on the application and estimation of the system behavior over the prediction horizon is using a performance model. Here optimization of the system behavior is carried out by taking into consideration the minimization of the cost incurred to the application.The advantage of such types of methods is that it can be applied over various performance management problems from systems with simple linear dynamics to systems with complex dynamics. The performance model can also be varied and affected with system dynamics as conditions in the environments like workload variation or errors in the system change.

Miyko Dori [9], describes about the memory reusing mechanism to decrease the amount of transferred data in a live migrating system. When we are considering the case of dynamic VM consolidation, virtual machines may migrate back to the host where it was once executed and so the memory image in that host can be reused, thus contributing to shorter migration time and greater optimizations by VM placement algorithms. In [10] it shows that this technique enables to reduce the total migration time. In this technique dirty pages alone need to be transferred to the former host.

The aim of load balancing in the cloud computing environment is to provide on demand resources with high availability. But often load balancing approaches suffer from various overheads. And they also fail to avoid deadlocks when there more requests competing for the same resource at the same time when the available resources are insufficient to service the arrived requests. The ELBA approach [11] using the efficient cloud management system helps to overcome the aforementioned limitations. This approach yields less response time compared to the existing approach. Less response time reduces job rejections and accelerates the business performance.

### III. RESOURCE OVERBOOKING

Resource overbooking is nothing but reserving resources in advance. In [13] it gives a detailed description of the overbooking technique and what are the advantages that the customer can benefit from this technique in a cloud. It is more useful in the concept of virtualization, clearly say virtual desktops.

Resource overbooking is the technique that can establish an increase of the average utilization of hosts in a data center by reserving fewer resources than required in worst case. Since more virtual desktops can be allocated to a host, the cost for the service provider related to investment in hardware equipment, server maintenance cost and energy cost can be reduced. The risk parameter that is limiting the degree of overbooking is the risk to affect the user satisfaction [14]. Fiedler in this proposed method suggests a careful overbooking for network virtualization and it also obeys service level agreements (SLA) for full and limited availability. Full availability means the availability of all there required resources for particular request.

Limited availability stands for the availability of certain share of required resources that are statically guaranteed at given degrees.

Urgaonkar et al. in [15] mentioned about how to maximize the revenue through overbooking. They suggest that provisioning cluster resources based on the worst case needs of the application results in low average utilization. it is because of the fact that average resource requirements of an application are normally smaller than its worst case requirements. And also the resources tend to idle when at times when the application does not utilize its peak reserved share. In [12] it summarizes that in shared hosting platforms techniques to overbook (i.e. under provision) resources in a guarded manner will outcome in revenue maximization through optimized usage

## IV. SYSTEM MODEL

In typical desktop-as-a-service architecture When a request to launch a remote desktop session arrives, in the site admission control system checks which servers have enough free resources to host the session. At runtime, a session admission control system checks if there are enough resources on the host to execute an additional application inside the remote desktop session. If not enough resources are available, the application is put in a queue and can only start when there are enough resources on the host. It consumes high investment cost at the service provider point of view and its also increases SLA violation.

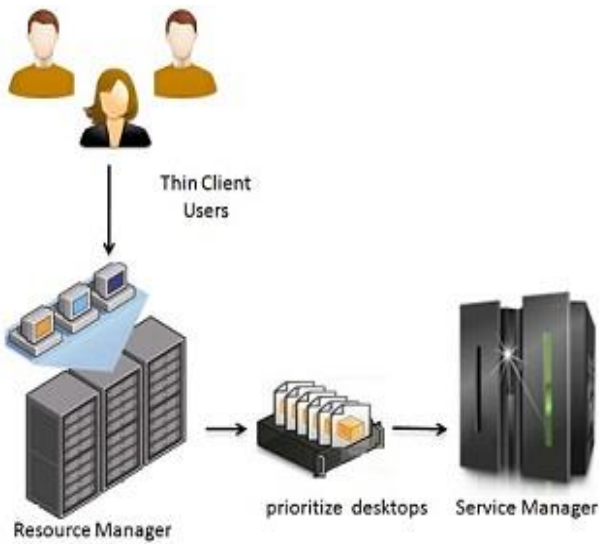

*Figure 2 .Intelligent Resource Allocation System Architecture*

We assume that there are M hosts in the datacenter and N users are subscribed to the desktop-as-a-service in the hybrid environment. All hosts are having the limited processing power, Based on their requirements the resources are allotted which will be less than their worst case requirement. This reservation of resources in advance can be named as Resource overbooking which will be discussed in the III section. The overall system model is given in figure 2

Here we introduce one new Resource Manager Component whose main function is to monitor client Request coming from various resources and execute them according to their priority. In IRA Technique we are giving 6 different level of priority to client request based on Requirement of various service level agreement (SLA) Factors which are elaborate in below table .

| PRIORITY | SLA FACTORS | | | | |
|---|---|---|---|---|---|
| | Throughput | Reliability | Durability | Agility | Security |
| Priority1 | yes | yes | yes | yes | yes |
| Priority2 | yes | no | yes | yes | yes |
| Priority3 | yes | no | no | yes | yes |
| Priority4 | yes | no | no | no | yes |
| Priority5 | yes | no | no | no | no |
| Priority6 | no | no | no | no | no |

*Table 1. Priority Table Based on SLA Factors*

- **Throughput:** System Response speed. How Quick System will be Response to Arriving particular Request .
- **Reliability:** System availability. All the servers on every continent are available? Or just one is available? It pays to define those definitions.
- **Durability:** How likely to lose data .Which guarantees that transactions that have committed will survive permanently.
- **Agility:** How quickly the provider responds to load changes. if datacenter is not able to execute particular request then how it will be react is shown by Agility.
- **Security:** A customer must know his security requirements and what controls and alliance patterns are essential to meet those requirements. A provider must comprehend what they must deliver to the customer to enable the appropriate controls and alliance patterns.

If any Client request Require all this SLA factors then it will be assign highest priority and it will be executed first. If any Client request Require lesser SLA Factors then it will be assign lowest priority and it will be executed later. Through this way we can reduces SLA Violation and batter User Experience, increase service provider`s profit margin.

## V. SIMULATION ENVIRONMENT

Our simulation atmosphere is an extension of CloudSim 3.0.3 toolkit [16]. There are some distinctive features for CloudSim, they are (1) accessibility of virtualization engine, which helps in creation and management of various, co-hosted and autonomous virtualized services on a data center node (2) better flexibility to change between time-shared and space-shared allocation of processing cores to virtualized services. These popular features of CloudSim would speed up the progress of new resource allocation schemes and scheduling algorithms for Cloud computing [17].

CloudSim acquire a realistic model to drift virtual desktops from one host to the other. And the phase duration depends on the time taken for migrating the assigned memory to the virtual desktop. Here a Rule-Based Resource Management system is implemented which will perform the VM scheduling according to the priority of the incoming requests.

## VI. EXPERIMENTAL EVOLUTION

The main functions that are implemented under the intelligent resource allocation (IRA) Technique are as follows.

- Every new request has to be assigned with six different set of priority i.e. highest priority to lowest priority according to their nature and SLA factors.

- IF the requests which are require all six SLA factors will be assign highest priority. And the other requests will be assign priority on basis of requirement their SLA factors.

- Perform resource overbooking by using the same set of rules performed for the previously implemented system [1].

- Then the resource manager has to analyze the incoming requests for their priority and which mode of operation they prefer.

- After monitor request priority resource manager send highest priority to lowest priority request to service manager for execution thus request having higher priority will be executed first and request having lowest priority will be executed later.

- Finally a system which is scheduled based on the priority of the requests is created and we have to prove that the average SLA violation imposed by the intelligent resource allocation (IRA) Technique is very less as compared to the previously implemented system [1].

We can also find that the resource utilization is high and the system can be extended for security concerns ,another advantage of IRA technique is it makes DaaS more reliable , durable , secure , and good response speed compared to the previously implemented system because of their SLA factors [1].

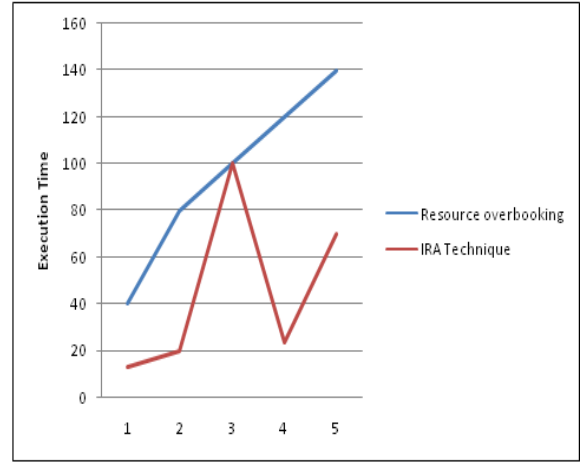

*Figure.3 Performance comparison between IRA(priority-based) Technique and the Resource overbooking strategy.*

Here a comparison is made between the system which is already implemented in [1] and the proposed the intelligent resource allocation (IRA) Technique . The results are shown in the graph in figure 3. It indicates that request 1 is getting highest priority, thus execution time is very low while request 3 is getting lowest priority thus its execution time is very high .

## VII. CONCLUSIONS

The concept of Desktop-as-a-Service in cloud computing, i.e. executing applications in virtual desktops on remote servers, is very interesting because it allow access to any type of application from any device.

Current desktop-as-a-service systems are mainly installed in LAN environments. Extending this to WAN environments involves important challenges to efficiently handle the typical large amount of geographically distributed and potential thin users. But It consumes high investment cost at the service provider point of view and its also increases SLA violation.

Intelligent Resource allocation technique provide six different level of priority for particular request based on requirement of SLA factors . through priority SLA violation we have to prove that the average SLA violation imposed by the intelligent resource allocation (IRA) Technique is very less as compared to the previously implemented system .We can also find that the resource utilization is high and the system can be more reliable , durable , secure , and good response speed compared to the previously implemented system because of their SLA factors .

## ACKNOWLEDGEMENT

I wish to thank, Dr. Ajay Shanker Singh and Prof. Rajanikanth for their valuable motivation, guidance and suggestion, which helped me for completion this Research paper.